\documentclass[sigconf]{acmart}

\AtBeginDocument{%
  \providecommand\BibTeX{{%
    \normalfont B\kern-0.5em{\scshape i\kern-0.25em b}\kern-0.8em\TeX}}}

\setcopyright{acmlicensed}
\copyrightyear{2024}
\acmYear{2024}
\acmDOI{XXXXXXX.XXXXXXX}

\acmConference[KDD]{ACM KDD 2024}{Aug 25--29,
  2024}{Barcelona, Spain}
\acmISBN{978-1-4503-XXXX-X/18/06}

\usepackage[ruled,vlined,linesnumbered,commentsnumbered]{algorithm2e}
\usepackage{amsmath}
\usepackage{caption}
\usepackage{subcaption}
\usepackage{hyperref}
\hypersetup{colorlinks=true,
    linkcolor=blue,
    filecolor=magenta,      
    urlcolor=cyan
}

\newcounter{def_count}
\newtheorem{definition}[def_count]{Definition}

\newenvironment{compact_enum}{
\begin{itemize}
  \setlength{\itemsep}{-0pt}
  \setlength{\parskip}{-0pt}
  \setlength{\parsep}{-0pt}
  \setlength{\itemindent}{-0pt}
  \labelsep=12pt
}{\end{itemize}}
\begin{document}

\title{Learning to Retrieve for Job Matching}


\author{{\small Jianqiang Shen$^+$, Yuchin Juan$^+$, Shaobo Zhang, Ping Liu, Wen Pu, Sriram Vasudevan, Qingquan Song\\ Fedor Borisyuk, Kay Qianqi Shen, Haichao Wei, Yunxiang Ren, Yeou S. Chiou, Sicong Kuang, Yuan Yin\\ Ben Zheng, Muchen Wu, Shaghayegh Gharghabi, Xiaoqing Wang, Huichao Xue, Qi Guo\\Daniel Hewlett, Luke Simon, Liangjie Hong, Wenjing Zhang}}
\affiliation{%
{\small
  \institution{LinkedIn Inc.}
  \country{}
}
}

\renewcommand{\shortauthors}{X and Y, et al.}

\begin{abstract}
\let\thefootnote\relax\footnote{~~~~ \\ $^+$ Equal contribution to this work. }
  Web-scale search systems typically tackle the scalability challenge with a two-step paradigm: retrieval and ranking. The retrieval step, also known as candidate selection, often involves extracting standardized entities, creating an inverted index, and performing term matching for retrieval. Such traditional methods require manual and time-consuming development of query models. In this paper, we discuss applying learning-to-retrieve technology to enhance LinkedIn's job search and recommendation systems.
In the realm of promoted jobs, the key objective is to improve the quality of applicants, thereby delivering value to recruiter customers. To achieve this, we leverage confirmed hire data to construct a graph that evaluates a seeker's qualification for a job, and utilize learned links for retrieval. 
Our learned model is easy to explain, debug and adjust.
On the other hand, the focus for organic jobs is to optimize seeker engagement. We accomplished this by training embeddings for personalized retrieval, fortified by a set of rules derived from the categorization of member feedbacks.
In addition to a solution based on a conventional inverted index, we developed an on-GPU solution capable of supporting both \textit{KNN} and term matching efficiently. 
\end{abstract}



\maketitle

\section{Introduction}
As one of the largest professional networking platforms globally, LinkedIn is a hub for job seekers and recruiters, with 65M+ job seekers utilizing the search and recommendation services weekly to discover millions of open job listings. To enable realtime personalization for job seekers, we adopted the classic two-stage paradigm of retrieval and ranking to tackle the scalability challenge. The retrieval layer, also known as candidate selection, chooses a small set of relevant jobs from the set of all jobs, after which the ranking layer performs a more computationally expensive second-pass scoring and sorting of the resulting candidate set. This paper focuses on improving the methodology and systems for retrieval. 

The inverted index is a core concept of traditional retrieval systems \cite{zobel2006inverted}. The process of building an inverted index involves converting documents into individual terms and recording the list of documents where a term appears, to enable efficient and rapid retrieval of documents containing specific terms. In Linkedin platform, a job posting is a document.
Manually crafted query models are typically used to define how member-provided keywords are to be translated into inverted index retrieval terms, with the overall goal of achieving good precision and recall while keeping the retrieved candidate set compact.
While prior efforts have attempted to use machine learning to refine query models \cite{broder2003efficient, borisyuk2016casmos}, expressing member intent precisely through keywords remains challenging. The advent of deep learning in recent times has resulted in embedding-based matching gaining popularity \cite{huang2020embedding, li2021embedding}. The idea here is to encode items into a lower-dimensional space, where vector proximity such as cosine similarity is used to measure semantic affinity. This approach excels in capturing nuanced relationships and semantic meanings, thereby enhancing accuracy in retrieving pertinent information.


At LinkedIn, job matching needs to be real-time due to the highly dynamic nature of the job ecosystem. Unlike conventional recommendation systems, job matching involves qualification constraints, demanding the ability to swiftly debug and explain.
We historically relied on extracted attributes for retrieval \cite{li2016get, li2020deep}. We first extract standardized entities (job titles, company names, required skills, geographic locations etc. from prebuilt taxonomies) from job descriptions and build an inverted index where the terms are these entities and the documents are the job postings. Online, we use the same extraction and standardization process to map a query to a known set of entities. Handcrafted clauses incorporating query expansion \cite{efthimiadis1996query} and rewriting operations \cite{papakonstantinou1999query} are then leveraged to match jobs. These approaches have the advantage of being able to explain retrieved results via matched attributes, and the result set can be easily refined by updating these attributes.

Manually crafting query models is time-consuming and sub-optimal, and the machine learning approaches to address this can be classified into two categories. The first involves expanding, pruning, or tweaking clauses to optimize results from the inverted index \cite{broder2003efficient, tonellotto2013efficient}. Recent efforts at LinkedIn \cite{borisyuk2016casmos, xue2020ranking} have taken this approach, a natural evolution of handcrafted query models to instead focus on learning attribute clauses. The second category involves extracting and utilizing semantic meanings from both queries and documents (often in an embedding space) to optimize the search process \cite{huang2020embedding, li2021embedding}.

In this paper, we formalize our learning-to-retrieve problem in both promoted and organic domains. In promoted channels, our goal is to dynamically regulate the equilibrium between job liquidity and qualification. This is achieved by constructing graphs that link seekers and jobs to help hirers target the right candidates. In organic channels, we optimize for member engagement through an Embedding Based Retrieval (EBR) system safeguarded by term matching rules. In both cases, we ensure that a seeker's qualification does not significantly deviate from the requirements of matched jobs. Among the jobs meeting these constraints, we retrieve the most engaging ones for late-state ranking. Finally, in addition to the more common approach of leveraging an inverted index, which often entails a time-consuming process of encoding, index construction, and deployment, we propose a novel exhaustive search system based on GPUs that supports KNN (K-Nearest Neighbors), thereby significantly improving relevance and productivity compared to its ANN (Approximate Nearest Neighbor) counterparts.

This paper makes several contributions. Firstly, we systematically study the problem of job matching, formulate its candidate generation problem and proposed multiple solutions. Secondly, we introduce a novel exhaustive search system based on GPUs, demonstrating superior performance compared to inverted index-based systems. Thirdly, we share practical learning lessons that can significantly benefit endeavors in large-scale learning to retrieve.

\begin{figure}[t] 
\centering
\includegraphics[width=2.5in]{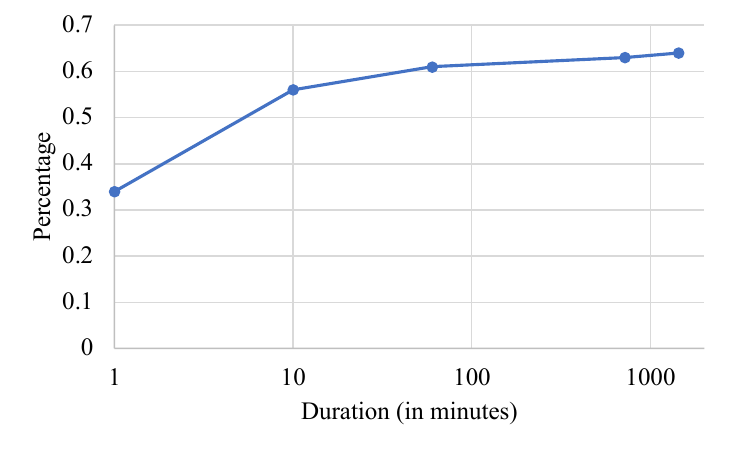}
\vspace{-12pt}
\caption{We computed the percentage of seekers who engage in at least two sessions within a specified duration per month.}
\label{fig_sessions}
\vspace{-4pt}
\end{figure}

\section{Problem Definition and Challenges}



\begin{figure}[t] 
\centering
\vspace{-6pt}
\includegraphics[width=3.0in]{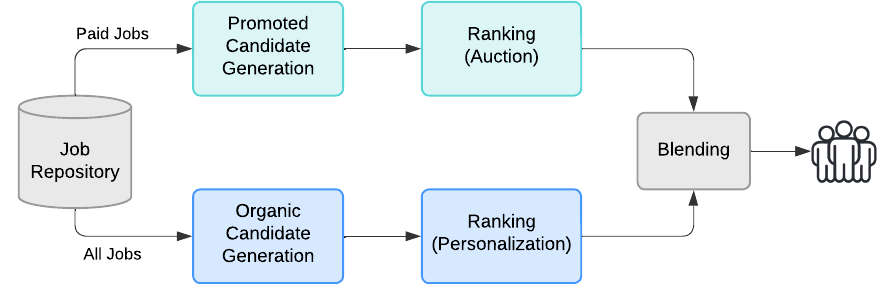}
\vspace{-10pt}
\caption{LinkedIn job matching system has a flow dedicated to promoted jobs, and a flow dedicated to organic content.}
\label{fig_overall_flow}
\vspace{-4pt}
\end{figure}

LinkedIn serves as the hub connecting 65M+ unique weekly active job seekers to economic opportunities, including tens of millions of free (oragnic) and paid (promoted) job listings. Our job ecosystem is highly dynamic, with the creation of numerous new jobs and the closure of older ones daily.
Job seekers are active on our platform and Figure~\ref{fig_sessions} shows that over 50\% of job seekers resume their job-seeking sessions within 10 minutes at least once per month, underscoring the importance of realtime signals to enhance the personalization experience.
Our job matching system employs the classic 2-stage paradigm, incorporating retrieval and ranking, to address scalability challenges. The system contains one ranking flow for promoted jobs and one ranking flow for all jobs, as shown in Figure~\ref{fig_overall_flow}.
In both flows, the retrieval phase constructs a query utilizing context information (member detail for recommendations, and member + search keywords for search), fetching candidates based on this query. The promoted flow then ranks the chosen subset of jobs based on auction, while the organic flow ranks the chosen jobs based on potential engagement, a.k.a personalization.
After being merged and processed by the blending model which balances the business objectives, the refined results are presented to the job seeker. 

There are distinct relevance requirements for our organic and promoted pipelines. In the organic pipeline, our primary objective is to foster growth and retention of job seekers, hence personalization to effectively present relevant opportunities is crucial. In the promoted pipeline, where charges are incurred for each click, delivering value to job posters becomes critical.
Employers want qualified candidates for their job openings, 
yet pinpointing the ideal audience can be more challenging than anticipated. Consider a scenario where a job posting aims to recruit a backend developer proficient in Java; however, the system may match frontend developers with Java expertise to this role.
This example shows that in addition to explicit facets, some ``targeting'' functionality is needed to ensure value delivery to job posters. Given a job $j$ and a seeker $s$, our learning-to-retrieve problem can be formalized as below:

\begin{definition} 
A {\normalfont Qualifier} indicator $\varrho(s, j)$ outputs 1 if seeker $s$ meets the basic qualification of job $j$, and 0 otherwise. An {\normalfont Engagement} indicator $\epsilon(s, j)$ outputs 1 if $s$ engages with $j$. In our application, we use `applies' as a metric to measure engagement.
\end{definition}

\begin{definition} 
For the {\normalfont promoted pipeline}, our goal is to learn a retrieval model $\Theta$ to select $k$ jobs out of a set of jobs $\mathcal{J}$, so that
\begin{align*}
\max_{\Theta }  \sum_{j\in {\Theta(\mathcal{J}, k)}}{\varrho(s, j)}
\end{align*} 
\label{promoted_learning}
\end{definition}

\begin{definition} 
For the {\normalfont organic pipeline}, our goal is to learn a retrieval model $\Phi$ to select $k$ jobs out of a set of jobs $\mathcal{J}$, so that
\begin{align*}
\max_{\Phi }  \sum_{j\in {\Phi(\mathcal{J}, k)}}{E(\epsilon(s, j))}
\end{align*} 
\label{organic_learning}
\end{definition}

Pre-processing or post-processing steps might be involved to handle stringent constraints, such as search facets. There is a key distinction between promoted and organic flows. In the promoted flow, we aim to efficiently target the right audiences based on their qualifications, to deliver value to those paid recruiters. Manual targeting may not consistently produce optimal results due to the intricacies of LinkedIn member profile data and the varying experience levels of recruiters. Our learning-to-retrieve approach focuses on building targeting rules for recruiters, with an emphasis on the capability to explain and manually adjust these learned rules.
In the organic flow, our goal is to present job seekers with positions aligned with their interests. This objective is achieved by analyzing their past activities and considering the actions of members whose profiles resemble theirs.
We will present the technical details in the following sections.

\section{Previous Work}

\textbf{Job recommendation} has some unique challenges compared with other recommendation problems \cite{de2021job}. To retain LinkedIn's valuable recruiter customers,  it is critical to deliver a sufficient number of applications from qualified candidates \cite{kenthapadi2017personalized}. 
It is not desirable either for the system to deliver too many applications to any posted jobs with one or a few openings, as the amount of effort for the job poster to interview would become much greater than expected. 
To achieve those goals, we need to understand unstructured job postings with noisy information. 
\cite{li2020deep} applied deep transfer learning to create domain-specific job
understanding models.  Jobs are represented by professional entities, including titles, skills, companies, and assessment questions.

In \cite{li2016get}, the authors explored three types of entity-aware features and incorporated them into the job search ranking function, and showed that  a new model with these features resulted in better user engagement. To enable semantic capabilities, the queries are segmented and mapped to entities from a known ontology, and the documents are also standardized by the same mechanism and indexed on these standardized entities. The authors therefore incorporate query-job matching features into the ranking model, making the model retrieval-aware and better aligning the two stages. 
\cite{lu2013recommender} used a directed, weighted, and multi-relational graph to model the job seeker activity, and rank items according to their relevance to the target user.
\cite{bian2020learning} propose a multi-view co-teaching network from sparse interaction data for job-resume matching, to mitigate the sparse and noisy issues in job-resume interaction data.

\textbf{Term based learning-to-retrieve} techniques~\cite{quinlan2014c4} require mapping terms to known ontologies and/or query expansion strategies ~\cite{li2014semantic}. They focus on evolving traditional inverted index based  systems to deliver enhanced candidate generation and improved personalization ~\cite{borisyuk2016casmos, xue2020ranking, li2016get}. Models to retrieve (query generation) and models to rank (document scoring) are based on the terms that the documents are indexed by. As such, the performance of these systems is highly dependent on the accuracy of term generation. At LinkedIn, member profiles and job postings go through standardizers that extract structured information like \textit{job title}, \textit{job company}, \textit{skills} and \textit{geographic location} and standardize them to a known set of entities, which are then used as terms to index the jobs. Misclassifications at this stage can result in reduced precision and recall of the overall search and recommendation system.

In~\cite{borisyuk2016casmos}, the authors propose performing candidate generation using only Weighted AND (WAND) queries~\cite{broder2003efficient}, with the potential clauses being all possible conjunctions of atomic attributes. A logistic regression model is trained using the conjunctive clauses as features, with modifications made to restrict the coefficients to non-negative values. While this work proposes a novel method to cast query generation as a machine learning problem, a major drawback is that the same WAND query is used for all users.
~\cite{xue2020ranking} takes a different approach, focusing on ranking user attributes and using a disjunction of the top $k$ clauses to retrieve jobs. This not only overcomes the limitation in~\cite{borisyuk2016casmos} of a single global query by enabling personalized query generation, but also tackles the training complexity and latency overheads seen in that work. It however achieves this efficiency by not using conjunctions or negations as clauses. Another limitation is that it does not handle inter-attribute similarity within the top $k$ chosen clauses.

\textbf{Embedding based retrieval (EBR)} has been successfully applied for years in web search engines, for image and media retrieval and simple text-only retrieval tasks~\cite{mitra2018introduction}. ~\cite{huang2020embedding} extends this idea to incorporate text, user, context and social graph information into a \textit{unified embedding} to retrieve documents that are not only relevant to the query but also personalized to the user. This work also proposes a hybrid framework to deploy at industrial scale alongside a more traditional Boolean retrieval system, to handle both approximate and exact match requirements. The inverted index system is augmented to support Approximate Nearest Neighbor (ANN) by adding new query operators that support radius-based retrieval. \textit{Coarse quantizations} of the embeddings are generated ~\cite{johnson2019billion} and used as the terms to index documents. 

In~\cite{li2021embedding}, the authors improve upon~\cite{huang2020embedding} by using a ``Multi-Granular Semantic Unit'' to discover the meaning of user queries and a ``User Behavior Attention Unit'' to capture user preferences. The former captures the semantics of a query at various levels such as character and word segment sequences and also uses historical queries as additional context. The latter combines real-time, short-term and long-term user-product interactions to improve personalization. The EBR solution was deployed in parallel with collaborative filtering and term-matching systems, and relevance control was enabled by using exact matches for navigational queries and using Boolean filters to make EBR results more precise.

\section{Graph for Auto Targeting}

\begin{figure}[!t] 
\centering
\includegraphics[width=1.8in]{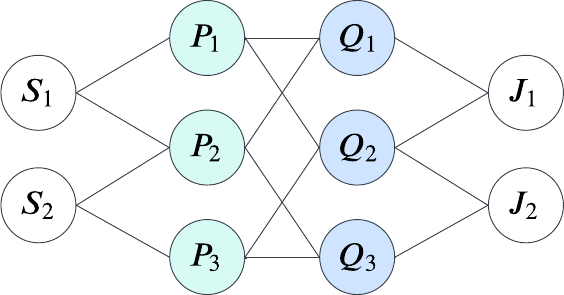}
\vspace{-6pt}
\caption{We map each seeker $S$ and each job $J$ to a segment $P$ or $Q$, and learn links between seeker and job segments.}
\label{fig_sorting_hat}
\vspace{-3pt}
\end{figure}

A primary goal within the promoted flow is to ensure applicant quality, thereby providing value to recruiter customers. Additionally, it is desirable to estimate and tune auction liquidity for both members and jobs to ensure that a job consistently receives an adequate number of applicants while managing its budget effectively. 
In this section, we elaborate on the construction of a graph-based framework for determining a candidate's qualification for a job. 

The key concept involves constructing a graph, as depicted in Figure~\ref{fig_sorting_hat}, to establish connections between members and jobs.
The model considers various attributes to construct segments, including member/job industry, titles, companies, skills, education, and the interactions between these attributes. 
Our objective is to establish a candidate selection mechanism that utilizes segments to bridge the gap between seekers and jobs. The segments abstraction simplifies the intricate matching logic between seekers and jobs, acting as implicit targeting facets.

\vspace{-2pt} 
\begin{algorithm}[t]
\DontPrintSemicolon
{\small
 \KwIn{$\mathcal{L}$: a set of link templates, $T$: training data, $\theta$: desired job liquidity per member.} 
 \KwOut{$\Omega$: links between seekers and jobs.}
  $\mathcal{D} \leftarrow \emptyset$  { //initialize the set of complex links and their stats}\; 
       
  \For{ each seeker \& job pair $\langle S_i, J_j \rangle \in T$}{
  	  $\bar{\mathcal{D}} \leftarrow$ all possible link combinations (complex links) based on $\mathcal{L}$\;
  	  $\mathcal{D} \leftarrow \mathcal{D} \cup \bar{\mathcal{D}}$ { //update the counting of each link comb} \;
  }
  \For{ each complex link $C \in \mathcal{D}$}{
  	  Discard $C$ from $\mathcal{D}$ {\bf if} $C$ does not have sufficient support\;
  	  $q_C \leftarrow$ compute a quality score for $C$ \;
  	  Discard $C$ from $\mathcal{D}$ {\bf if} $q_C$ is low\;  	  
  }
  \For{ each job seeker $S_i$}{
  		${\bar \Omega_i} \leftarrow$ all links in $\mathcal{D}$ that are compatible with $S_i$ \;
  		Sort links in ${\bar \Omega_i}$ based on their quality scores \;
  		$\Omega_i \leftarrow \emptyset$  { //initialize the final set of links for $S_i$}\; 
  	    \While{(liquidity of jobs associated with $\Omega_i) < \theta$ \& ${\bar \Omega_i} \neq \emptyset$ }{        
 			Pop link $L$ with the highest quality score from ${\bar \Omega_i}$ \;
 			$\Omega_i \leftarrow \Omega_i \cup L$ \;
	   }
  }  
  \Return $\Omega$ \;
}  
 \caption{Learn Links between Seekers and Jobs}
 \label{alg_link}
\end{algorithm}

\subsection{Learning Links between Seekers and Jobs}

For our monetization product, it is important to ensure that every job receives ample opportunities to participate in auctions, with payment only incurred for qualified candidates. We accomplish this by constructing graphs derived from \textit{confirmed hire} data and utilizing them for retrieval. In simpler terms, a \textit{confirmed hire} refers to a member who moved to a new company by applying through a LinkedIn job in our context. We aim to establish connections between a member's profile and their new job by analyzing attribute values. Such connections will then be utilized to target the right seekers during the retrieval phase. 
This ensures simplicity in explanation and allows for easy adjustments. 
Manually creating these connections is challenging, given the multitude of attributes associated with both seekers and jobs, some of which have high cardinality. For instance, the attribute ``skill,'' crucial to
both jobs and members, has over 335k+ unique values.

We leverage LinkedIn's knowledge graph on jobs and members to tackle this challenge. Each member is linked to a set of nodes representing their attribute values, as is each job. We endeavor to establish connections between the attribute value nodes of jobs and members using prior knowledge and confirmed hire data. We then extract common patterns from the connections. Let's start with the following definitions:

\begin{definition} 
A {\normalfont meta link} is a connection between a seeker attribute value and a job posting attribute value, and a {\normalfont complex link} (or simply, a {\normalfont link}) is a connection between a set of seeker attribute values and a set of job posting attribute values. A set could contain $\geq 1$ element.
\label{link_definitions}
\end{definition}

The idea is to create segments in the graph for seekers or jobs based on their attribute values, and then build links between these segments using confirmed hire data. The technique is outlined in Algorithm~\ref{alg_link}. 
We first leverage human knowledge to reduce the search space by defining a set of \textit{meta-link templates}. A \textit{template} outlines the permissible connections between member-side attributes and job-side attributes. An example is $\langle memberTitle \leftrightarrow jobTitle \rangle$, or $\langle memberSeniority \leftrightarrow jobSeniority \rangle$. From the confirmed hire data, we then enumerate all potential meta links between seekers and jobs that can align with templates. Using the above 2 templates as example, they could lead to 2 potential meta link candidates between seeker $S_i$ and job $J_j$: 
\begin{gather*}
\langle memberTitle = \text{ML Engineer} \leftrightarrow jobTitle = \text{NLP Engineer} \rangle \\
\langle memberSeniority = \text{Intern} \leftrightarrow jobSeniority = \text{Entry} \rangle 
\end{gather*} 
Assessing whether a seeker is a fit for a job solely based on a single meta link may not be reliable. Usually, multiple meta links are required to make a comprehensive decision. In the given example, a decision could be made by considering both the title and seniority. We achieve this by learning links, which in essence are an ensemble of multiple meta links.
At the same time, certain types of jobs may not necessitate a highly detailed matching logic. For instance, a member with the title ``Quality Control Microbiologist'' might be a strong match for a job with the title ``Drug Safety Associate,'' without requiring additional criteria such as seniority or skills. 
The configuration of meta-link ensembles should be adaptable -- some member-job pairings may require intricate meta-link ensembles, while others might suffice with just one or two meta links.

To create the link candidates, we generate all possible meta-link combinations for each confirmed hire. For a seeker and job pair with $k$ meta links, this would generate $2^k - 1$ link candidates. We then aggregate all link candidates from the entire confirmed hire data and cut off the less relevant long tail. In our setup, we filtered out link candidates if their frequency is less than 3 or if they result in more ``no hire'' instances than actual hires in the training data. This helped reduce the size of link candidates from 25M to 5M.

We compute a quality score for each link candidate, serving two purposes. First, these scores aid in additional graph pruning. Additionally, quality scores can be used to dynamically adjust job liquidity for each seeker. A straightforward method to calculate the quality score is to utilize the \textit{hire} to \textit{no hire} ratio associated with the link-candidate-connected seeker-job pairs. In this paper, we employ logistic regression with L1 regularization to train and predict the likelihood of a seeker and job pair being a confirmed hire. Each link candidate serves as an input feature to the model, and the learned weight is utilized as the quality score. We retain only those links with quality scores larger than a specified threshold, and the application of L1 regularization aids in efficiently pruning links.

\subsection{Online Targeting Using Links} \label{sec:implementation_SortingHat}

\begin{figure}[!t] 
\centering
\includegraphics[width=2.3in]{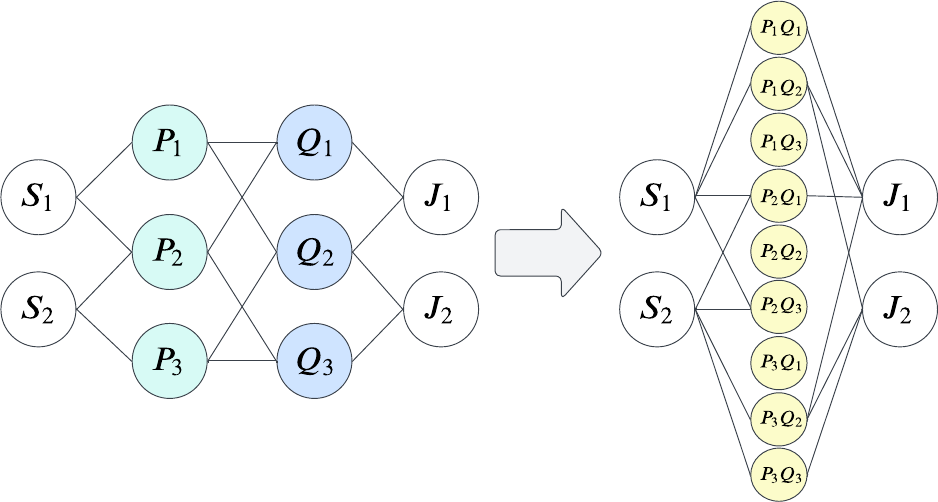}
\vspace{-6pt}
\caption{We transform a graph to 3 layers for online serving by replacing each seeker and job segment link with a node.}
\label{fig_collapse_graph}
\vspace{-7pt}
\end{figure}

Through the preceding steps, we essentially generate two types of segments -- one for seekers and one for jobs. Each segment is defined by a set of attribute values, and each job or member is mapped to one or multiple segments. When a link exists between a seeker segment and a job segment, it signifies that seekers associated with the seeker segment could be considered as suitable fits for the jobs linked to the job segment. The graph is easily understandable for humans, allowing us to make adjustments to the links when needed.

To deliver the best member experience, we want to ensure that each job seeker has access to a sufficient number of high-quality job opportunities. 
To achieve this, we fine-tune the graph for each seeker, dynamically adjusting the links to strike a balance between job liquidity and quality. In practice, this is achieved by pruning the mapping between seekers and seeker segments. 
Specifically, for a given seeker $S_i$, we remove all mappings between $S_i$ and seeker segments, treating them as candidate mappings. Each mapping is assigned a score, calculated as the average of the quality scores of the links associated with this seeker segment. Starting with the mapping having the highest score, we progressively reintroduce these mappings one by one until the job liquidity connected to $S_i$ meets the desired threshold or all mappings have been reintroduced. Please note that pruning can also be done on the job side. We focus on the seeker side to ensure the right member experience.

The refined graph is utilized for retrieval through an inverted index system. We simplify the graph by collapsing it from 4 layers to 3 layers, as depicted Figure~\ref{fig_collapse_graph}. Essentially, we combine the seeker segment layer and the job segment layer into a single layer. In this new layer, each node represents a pair of job segment and seeker segment. For each seeker, we populate its directly associated nodes into the key-value store, facilitating efficient lookups by seeker ID when requests are made. Similarly, for each job, we create an attribute for every node linked to that job and update the inverted index. The retrieval of a job for a seeker is contingent upon the presence of matching node values in both their profiles.

\section{Embedding for Personalization}

EBR has gained widespread popularity for its utilization of learned representations in retrieval tasks, demonstrating success in various domains, including YouTube recommendations \cite{yi2019sampling} and Facebook Search \cite{huang2020embedding}, and others. Unleashing the power of deep learning and semantic understanding in the retrieval layers will help job seekers find jobs more engaging to them. Given the  mentioned job qualification requirements, we introduced some straightforward rules to prevent completely off-target results.

\subsection{Objectives and Labeling}

For organic pipelines, our goal is to maximize seeker engagements. The optimization challenge lies in maximizing the similarity between positive seeker and job pairs, disregarding negative seeker and job pairs. We leverage member-engaged jobs from logged impression data and employ the following comprehensive strategies to generate negative labels for contrastive learning.

\textbf{In-Batch Negative Sampling} involves framing the problem as a multi-class classification, aiming to maximize the score of positive records while minimizing the scores of negative ones. The training data is derived from the logged engagement data, such as clicks or applications, on jobs in LinkedIn. Unlike conventional classification training data, this dataset exclusively comprises positive labels. In each batch, when considering a member $S_i$, the probability of member \( S_i \) clicking on job \( J_j \) is given by the softmax function:

\begin{equation}
P(J_j \mid S_i) = \frac{e^{z_{ij}}}{\sum_{k=1}^{B} e^{z_{ik}}}
\end{equation}

where \( B \) represents the total number of jobs in a batch, and \( z_{ij} \) represents the similarity score of job \( J_j \) for member \( S_i \) as derived from the model.

The loss function \( \mathcal{L} \) is defined as follows:

\begin{equation}
\mathcal{L} = -\frac{1}{N} \sum_{i=1}^{N} y_{ij} \log(P(J_j \mid S_i))
\end{equation}

where \( N \) represents the total number of samples. Here, \( y_{ij} \) serves as an indicator function, taking a value of 1 if member \( S_i \) clicked on job \( J_j \), and 0 otherwise. The straightforward interpretation of optimizing the corresponding softmax loss entails utilizing jobs that have garnered interest from other members as negative samples for the current member in the context of this multi-class ``click prediction'' classification task.

\textbf{Random Easy Negative Sampling} from the inventory has proven beneficial in addressing bias \cite{yang2020mixed}. While being easy to implement, In-Batch Negative Sampling introduces selection bias by excluding items with no user feedback, whereas Random Easy Negative Sampling directly samples some items as negative samples from the inventory
In-Batch Negative Sampling and Random Easy Negative Sampling can be combined to enhance training data. A straightforward implementation begins by directly sampling from indexed jobs. Assuming a batch of training data contains $m$ rows of member-job pairs, a training matrix based on In-Batch Negative Sampling is generated by multiplying two vectors to get a $m \times m$ matrix. Subsequently, $n$ jobs are sampled from the inventory, distributed across $p$ batches. In each batch, $n/p$  jobs are appended to the job vector $J$, resulting in a  $m \times d$ matrix where  $d =m+ n/p$. So it is still a multi-class classification task but the negatives are not only from the batch but also from inventory.

\textbf{Online Hard Negative Sampling} proves advantageous to fine-tune models by emphasizing hard negative samples -- those instances where the model makes incorrect predictions, as highlighted in \cite{huang2020embedding}. This refinement is particularly beneficial after the model has been adequately initialized. More specifically, within each batch, an effective strategy for sampling hard negative instances involves leveraging the model itself to rank them, retaining only the top $k$ samples (excluding the positives) as negatives. In the context of batch softmax loss, this leads to a reduction in the training matrix size from $m\times d$ to $m\times k$. 

\subsection{Modeling Strategy}

\begin{figure}[!t] 
\centering
\includegraphics[width=3.3in]{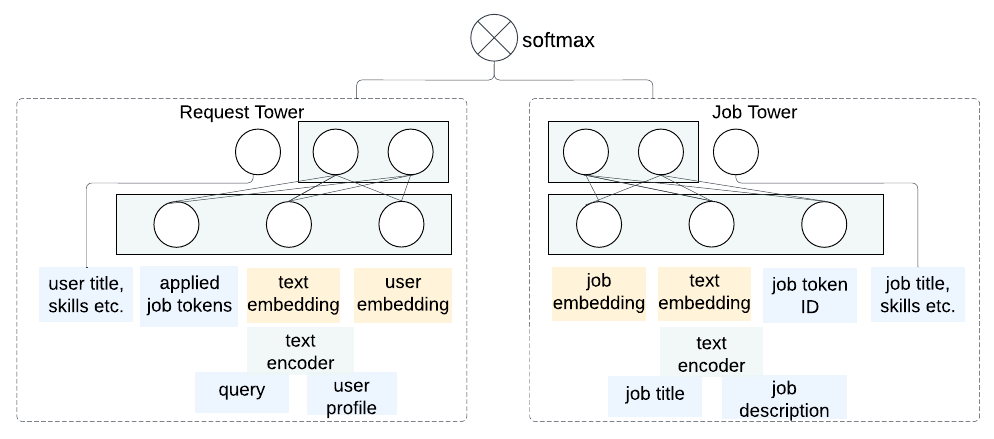}
\vspace{-6pt}
\caption{The two-tower model architecture used in our EBR.}
\label{fig_ebr_arch}
\vspace{-8pt}
\end{figure}

In parallel with other EBR projects, we have embraced a wide \& deep two-tower model architecture. One tower is dedicated to generating job embeddings, while the other focuses on generating request embeddings. The two-tower DNN model learns representations from request and job sides, along with a function to measure the similarity between requests and jobs, illustrated in Figure~\ref{fig_ebr_arch}. In the case of recommendation, the request embedding primarily originates from the seeker's profile and their past engagements on LinkedIn. For search, the approach is similar but also includes the query typed by the seeker. 
The raw text, including query, seeker profile, job title, description, is encoded into embeddings using a text encoder. These embeddings are then concatenated with other features, such as pretrained embeddings and member/job entity features (such as title, skill, seniority), along with engagement features on the seeker side.

Drawing inspiration from the approach outlined in \cite{liu2021que2search}, we leverage curriculum learning to refine the decision boundary of the EBR model via learning from harder negatives progressively. The EBR models were initially trained on data with easy negative samples in the first stage to warm up and obtain a coarse ranking ability among the entire corpus. Subsequently,  they were fine-tuned in a second stage using online hard negative samples filtering based on the similarity score within each batch. To be more specific, we fixed the batch size to be 2,048 in both training stages. In the first stage, each positive sample is associated with 2047 negative samples, encompassing a mix of in-batch negative samples and easy negative samples for computing the training loss. In the second stage, each positive sample is coupled with only the top $K$ negative samples selected based on the highest EBR prediction scores (cosine similarity in our experiments) for loss computation.

EBR models capture semantic relationships better than attribute matching and can improve relevance through better personalization. However, they can also lead to some poor results due to the lack of exact matching. Boolean rules can be used as a general relevance control mechanism, in addition to powering specific usecases like navigational searches \cite{li2021embedding}.
We learned this firsthand when we observed regressions in search and recommendation quality when we deployed EBR. To tackle this challenge, we developed a simple set of rules, by categorizing member feedback and distilling these failure modes into rules that could be enforced by traditional term-based matching. This rule-based quality control not only allows us to better explain results but also provides a principled approach to curating rules when addressing seeker feedback.

\subsection{System Implementation} \label{sec:implementation_GPU}

As previously discussed, our retrieval system must support hybrid retrieval expressions, with combinations of embeddings and terms. \textit{Term-based retrieval} (\textit{TBR}) is employed for diverse hard constraints, including search facets and rules designed to ensure quality safeguards. 
In terms of implementation, there are two approaches to EBR: \textit{fully-scanned EBR} (\textit{F-EBR}) and \textit{heuristic EBR} (\textit{H-EBR}). In F-EBR, the system exhaustively computes embedding scores against each document during retrieval. On the other hand, H-EBR employs heuristic information to guide computations and scores only a subset of documents, often using specific algorithms to partition embeddings, such as IVFPQ \cite{jegou2010product} and HNSW \cite{malkov2018efficient}.
Similarly, TBR can be categorized as either fully-scanned (F-TBR) or heuristic (H-TBR). F-TBR checks documents individually against predefined criteria, while H-TBR only examines a subset of documents, often relying on an inverted index.
In crafting our hybrid system, we paid close attention to the following design choices.

\textbf{F-EBR vs H-EBR.} H-EBR was primarily devised to address speed challenges by scanning only a small number of documents. However, implementing H-EBR introduces several additional challenges. Partition strategies like IVFPQ \cite{jegou2010product} or HNSW \cite{malkov2018efficient} require construction of additional models, thereby adding extra costs. A careful infrastructure design is needed to ensure that each embedding version aligns with the correct version of heuristics. This adds complexity to the system and makes operation more challenging. The performance bottleneck of these heuristics may diminish the relevance improvement of embeddings, complicating the task of driving business metrics. Further more, the speed advantage of H-EBR over F-EBR decreases when the TBR pass-rate is low, and more documents need to be returned from EBR.

\textbf{F-TBR vs H-TBR.} TBR typically relies on an inverted index, performing well given a low matching rate between queries and documents but exhibiting reduced efficiency with high matching rates. For instance, when retrieving documents matching ``title=SWE AND zipCode=95035'', the system efficiently finds the intersection between 2 sets: documents in zipcode 95035 and those with the title SWE. The time complexity of this is decided by the size of the smaller set. For queries like ``country=US AND title=Engineer'' where both sets are large,  computation becomes intensive. Heuristic rules such as \textit{early stopping} are often applied. For example, if there are 3M documents matching the query, the system may stop after finding 100K documents. However, this mechanism can be biased, leading to the potential exclusion of high-quality documents.

\textbf{EBR first or TBR first.} In addition to parallel execution and computing intersections, there are two strategies for efficiently integrating EBR and TBR to save computation. The first approach runs TBR first, and EBR is then performed among the documents selected by TBR. Instead, EBR can be executed first, assuming that returned documents will be sorted from the highest score to the lowest. TBR then performs ``post filtering'' until the target documents to return are satisfied. The decision on whether to run TBR first depends on the relative speed of TBR and EBR. If TBR is faster, running TBR first will be more efficient. In cases where the query includes both simple and complex boolean conditions, employing a mixed approach -- TBR using simple conditions, then EBR, and finally TBR using complex conditions could  improve efficiency.

\textbf{Communication cost.} Handling messages between TBR and EBR is critical to  hybrid systems. When TBR is run first, and the matching rate is high, we might need to pass massive data to the EBR stage. On the other hand, if EBR is run first and fails to pass sufficient candidates (due to low matchings from TBR), then we might need to call EBR again to retrieve more candidates. In both cases, efficient communication between the two stages becomes crucial. Given the stringent latency requirements of retrieval, deploying TBR and EBR in separate data centers or different machines is impractical due to communication costs. Even if TBR and EBR operate as independent programs on the same machine, the communication overhead at the OS level may still be excessively costly.

\begin{figure}[!t] 
\centering
\includegraphics[width=2.8in]{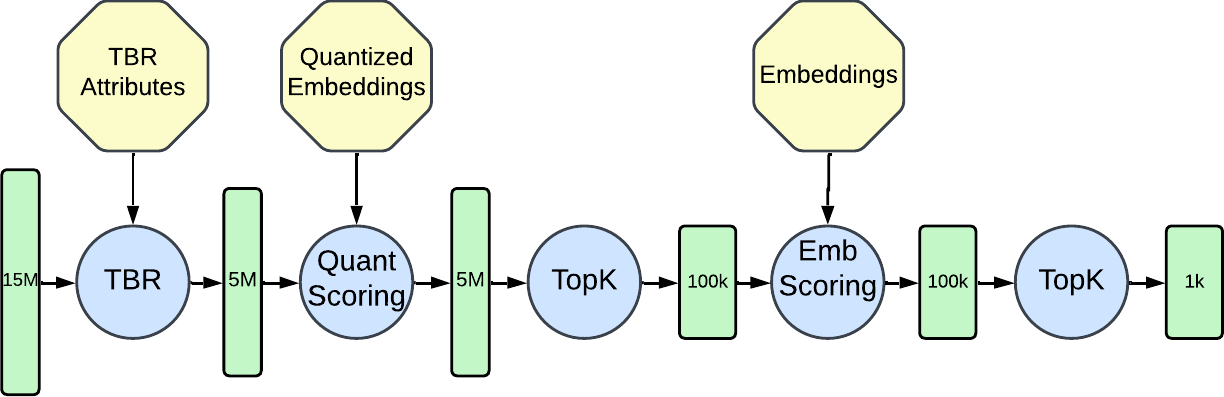}
\vspace{-6pt}
\caption{Illustrative flow of our on-GPU hybrid retrieval.}
\label{fig_on_gpu_sys}
\vspace{-4pt}
\end{figure}

Full scans on a large repository using CPUs  are in general impractical for retrieval. Traditional systems either runs H-TBR or H-EBR first to reduce the candidate size. As discussed, H-TBR is suboptimal with high matching rates, and H-EBR has the risk of losing the relevance gain from embedding improvements along with the challenge of extra operational costs.
Fortunately, with the computational power and parallelism offered by contemporary GPUs, implementing F-TBR and F-EBR systems becomes feasible. As we will show in the experiments, performing full scans on tens of millions of documents can be achieved in a few milliseconds. We have successfully implemented an efficient retrieval system entirely on GPUs. 
Different from other industry-deployed EBR systems that rely on inverted index and ANN \cite{li2021embedding, huang2020embedding}, our system operates on full scans and employs KNN based on matrices, executing F-TBR followed by F-EBR as shown in Figure~\ref{fig_on_gpu_sys}.
F-TBR and F-EBR are integrated in the same program \footnote{The code will be open sourced later on.}, eliminating communication costs between them. 

The system contains two matrices: a sparse matrix, where each row corresponds to a job posting and each column represents an attribute, and a dense matrix, with each row corresponding to an embedding vector representing a job posting. Each TBR query is transformed into Conjunctive Normal Form, where each clause is multiplied against the sparse matrix in parallel. For Embedding-Based Retrieval (EBR), each query is converted into a multiplication of the query embedding against the dense matrix, resulting in a vector with scores for each job. 
We optimized operations by harnessing the parallel and batching capabilities of GPUs.
To select the top $K$ from the result vector which has high dimension, instead of the conventional sorting method, we implemented a selection algorithm inspired by Bucket Sorting~\cite{cormen2022introduction}, with 5x improvement in efficiency based on benchmarks.

To further enhance efficiency, we implemented an optional coarse scoring step which allows to perform pre-selection using a quantization technique akin to ``OP+ORP''~\cite{li2023oporp} before EBR. For the embedding matrix of jobs that pass TBR, we do a random permutation on the matrix. Then we divide the permuted embedding matrix into $k$ equal-sized bins. In each bin, we apply a multiplication to each entry with a random sign and aggregate all the entries. This process generates $k$ samples for each embedding vector, facilitating quick similarity estimation for the selection of top candidates to pass to EBR. This approach is simple to implement and eliminates the need for prior clustering and centroid storage.

\section{Experimental Results}

We conducted evaluations and deployed the new learning-to-retrieve systems to the product for both organic and promoted flows. Compared with the existing system, which focuses on learning atomic member attributes for retrieval \cite{xue2020ranking}, our new systems show substantial improvements in key business metrics.

\subsection{Promoted Pipeline}

\begin{figure}[!t] 
\centering
\includegraphics[width=2.8in]{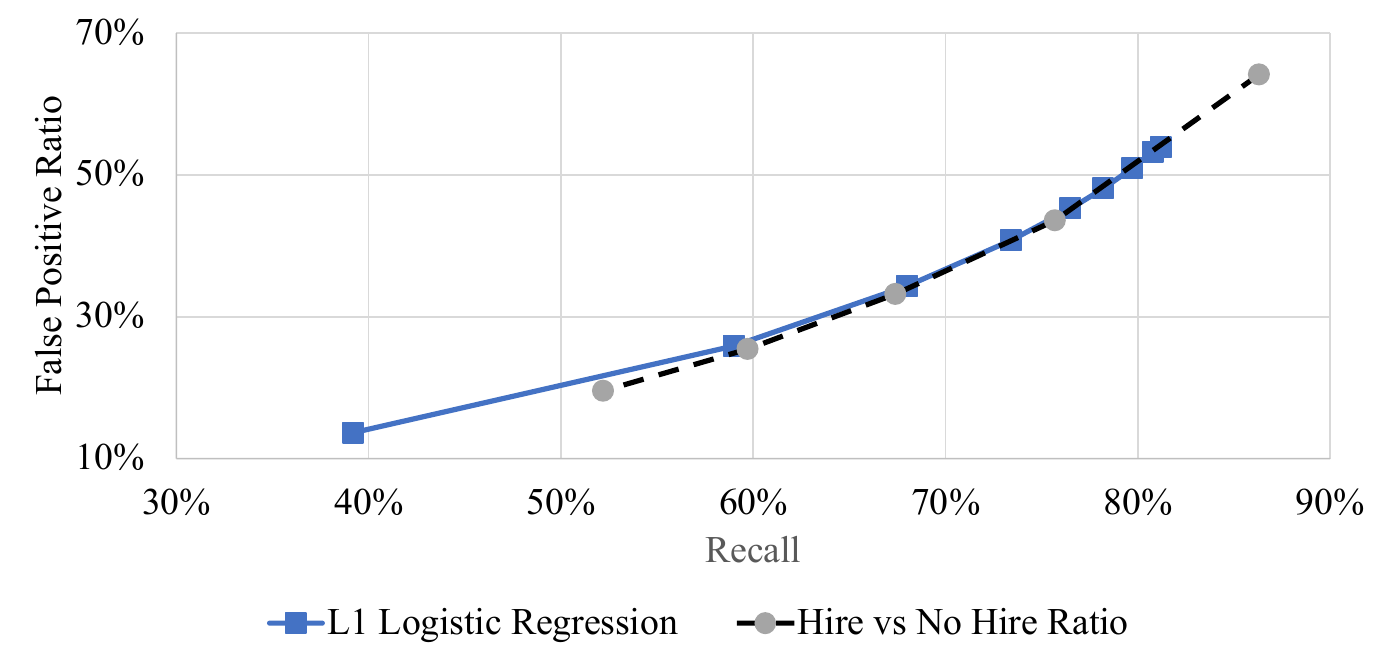}
\vspace{-8pt}
\caption{We plot the recall and false positive rates of 2 algorithms by varying the quality thresholds.}
\label{fig_sorting_hat_offline}
\vspace{-8pt}
\end{figure}

\textbf{Offline Results.} We assessed the ability of Algorithm~\ref{alg_link} to learn links using one month of confirmed hire data, comprising 457K records. We compared two methods to estimate link quality and perform pruning: one based on the ``hire'' to ``no hire'' ratio, and the other utilizing L1 logistic regression. By using various pruning thresholds, we obtained two metrics: the first is recall, measuring the percentage of job and seeker pairs that resulted in confirmed hires being linked. The second is the false positive ratio, measuring the percentage of jobs and seekers that are linked but did not lead to confirmed hires.

As shown in Figure~\ref{fig_sorting_hat_offline}, both algorithms have comparable false positive rates given the same recalls. Notably, L1 logistic regression, despite its slightly lower performance, demonstrates a considerable advantage by learning much sparser links -- on average, its link size is 77 times smaller. It learned link size ranges from 70K to 9K. This makes it more suitable for online serving.

\textbf{Online Results.} We conducted an online budget split A/B test \cite{website_bs}. Compared with the baseline model \cite{xue2020ranking}, our auto-targeting solution, leveraging learned links, resulted in an approximate $15\%$ relative increase in budget utilization. Given engagement metrics remained neutral, this now serves as our candidate generation solution for the promotion channel.

\subsection{Organic Pipeline}

\textbf{Offline Results.} Assessing the quality of embeddings is important when evaluating EBR. It serves as a foundational step to gauge the effectiveness of our modeling techniques, labels, features, and changes in model architectures. We define a general recall term:
\begin{equation}
 {recall}@k = \frac{1}{N}{\sum_{i=1}^{N} \frac{ \vert R_i \cap A_i \vert}{\vert A_i \vert}} 
\end{equation}

where $R_i$ represents the set of retrieved items for the $i$-th query and $A_i$ represents the set of actual relevant items.
We introduce two embedding quality recalls: ``In-Batch Recall'' and ``KNN recall.'' In-batch recall (noted as in-batch-$B$@$k$, $B$ is batch size) acts as a baseline metric, where the positives, indicating a click or job application within one batch, becomes the single target ID. The parameter $k$ can be tuned, and the retrieval set comprises the top $k$ jobs ranked by the EBR model within that batch. 
Using a batch to simulate the entire inventory may not be an ideal representation, hence we propose a second approach -- KNN recall (noted as offline-KNN@$k$). The KNN recall closely aligns with real production scenarios. It involves selecting the top $k$ results per request based on embedding models from the inventory, with the parameter $k$ tuned to match the number of jobs typically matched in production. 

\begin{table}[tb]
\small
\caption{Recall metrics for different learning strategies.}
\begin{tabular}{l|c}
\hline
variant & KNN@6400  \\ \hline \hline
Baseline (single stage with In-Batch Negative) & 0.7522  \\ \hline
Baseline + Easy Negative & 0.8132  \\ \hline
Baseline + Easy Negative + Curriculum Learning & 0.8164  \\ \hline 
\end{tabular}
\vspace{-2pt}
\label{tab:recall_labels}
\end{table}

The training and validation data is sourced from a month's worth of logged member activities on LinkedIn's job product channels. Specifically, the initial 30 days are allocated for training, while the last day is reserved for validation. The dataset has 135M clicks, 12M applies, and 3M job saves, totaling 150M records for training. Validation data has similar distribution, with 5M data points. 

The inclusion of negative samples plays a crucial role in developing a robust EBR model. Since majority candidates in the job marketplace retrieval corpus are negatives for a member, we introduce random easy negative samples into the data as a complement of original in-batch negatives, aiming to better mimic the candidate distribution within the full corpus. By fixing the training batch size to be $2048$ and exploring different easy negative ratios, we found that an introduction of 40 - 60\% easy negatives gives decent improvement on the final recall, as shown in Table~\ref{tab:recall_labels}. Curriculum Learning using top-1024 hardest negatives further improves the EBR performance.

\begin{figure}[!t] 
\centering
\includegraphics[width=2.5in]{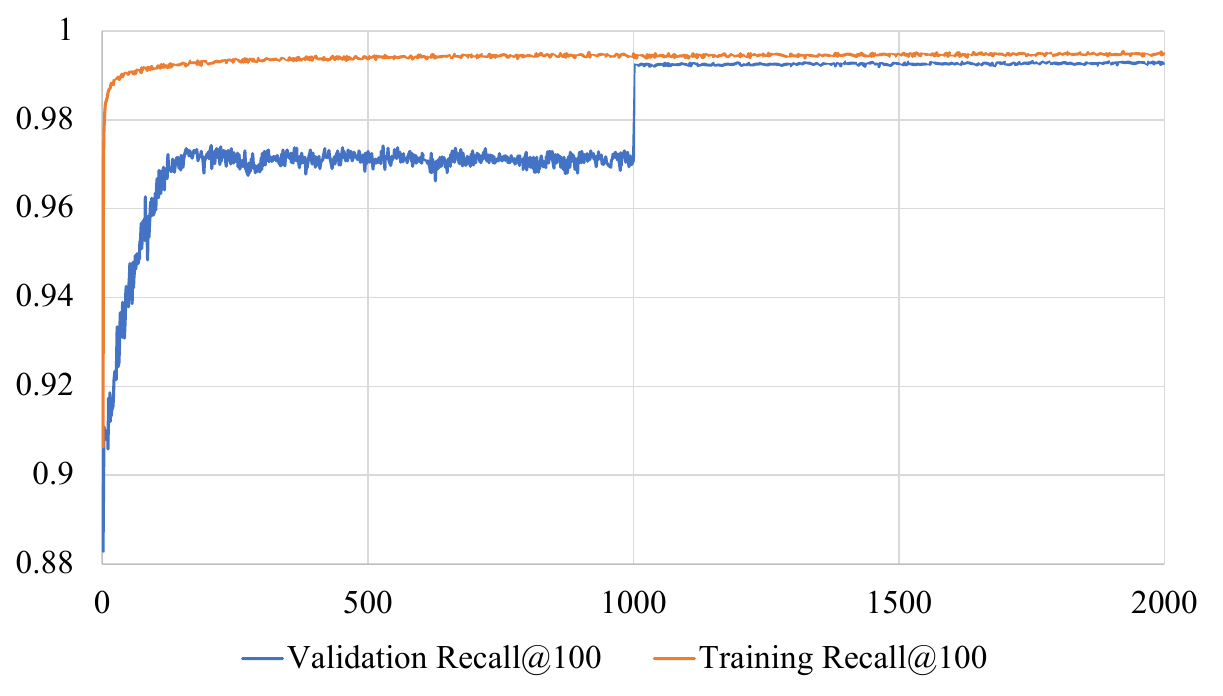}
\vspace{-4pt}
\caption{Curriculum learning curve with top-half hardest negative samples in the second stage.}
\label{fig_ebr_curve}
\vspace{-2pt}
\end{figure}

Our initial attempts on the second stage of curriculum learning with $K = 1$ resulted in catastrophic forgetting issues. We attribute this to the fact that learned boundaries are overfitted and biased to hard negatives thus losing the generalizability on the overall candidate corpus. To mitigate this issue, we enlarged the $K$ value and settled on $K = 1024$. We also found that introducing the elastic weight consolidation regularization \cite{kirkpatrick2017overcoming} and reducing the learning rate helped mitigate the issue orthogonally. These adjustments led to a notable achievement of $10\%$ in-batch-2048@10 and $2\%$ in-batch-2048@100. Figure~\ref{fig_ebr_curve} depicts the learning curve of the updated model, where the first 1k steps correspond to the first stage, and subsequent steps represent the second stage with an apparent bump on validation recall.

\begin{figure}
\vspace{-0pt}
\centering
\begin{subfigure}{.25\textwidth}
    \includegraphics[width=\textwidth]{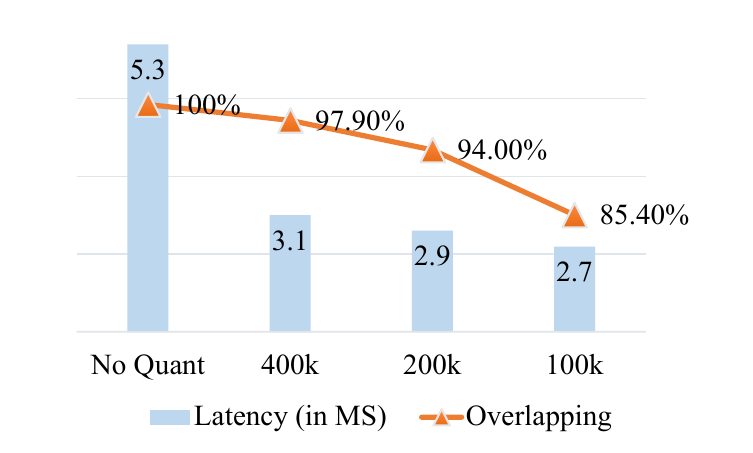}
    \captionsetup{justification=centering}
    \vspace{-6pt}
    \hspace{-22pt}    
    \caption{TBR matching rate = 33\%}
    \vspace{-12pt}
    \label{fig:perf_matching33}
\end{subfigure}%
\begin{subfigure}{.25\textwidth}
    \hspace{-10pt}
    \includegraphics[width=\textwidth]{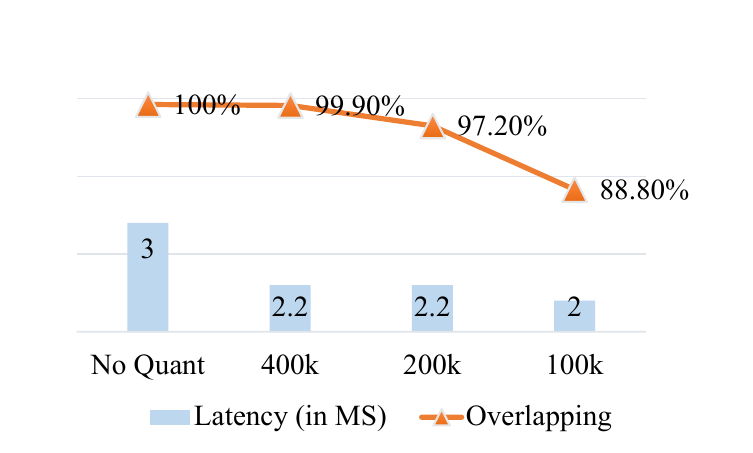}
    \captionsetup{justification=centering}
    \vspace{-6pt} 
    \caption{TBR matching rate = 10\%}
    \vspace{-12pt}
    \label{fig:perf_matching10}
\end{subfigure}
\caption{Benchmarks of On-GPU EBR given TBR matchings.}
\vspace{-6pt}
\label{fig:system_perf}
\end{figure}

\textbf{System Benchmarks.} We implemented our hybrid TBR + EBR system on GPUs using CUDA. To evaluate its performance, we conducted benchmarks by selecting the top 1K documents from a pool of 15M documents using an A100 GPU. Figure~\ref{fig:system_perf} shows the response time given various pass rates from TBR. We also compared these results with those obtained without quantization and presented the overlapping outcomes. Our system has low latency and effectively preserves the top results, with an ease of operation and maintenance.

\textbf{Online Results.} We deployed EBR in both job recommendation and search products, with the relevance results safeguarded by a simple set of rules derived from categorizing member feedbacks (including occupation and skill matching). 
It has yielded significant business metric improvements. 
Notably, in search, there has been a substantial relative increase in Successful Job Search Sessions, rising by $+2.37\%$. Additionally, job applications had a boost with $+1.45\%$ relative, while click rate had a positive relative uptick of $+1.49\%$. We saw similar engagement improvements in the recommendation product too. 

\section{Conclusions and Future Work}
We formalized the learning-to-retrieve problem for both promoted and organic domains, with a focus on dynamically regulating job liquidity and qualification in promoted channels and optimizing member engagement in organic channels. We ensured that a seeker's qualification aligns with matched job requirements to retrieve the most engaging jobs meeting constraints. This was achieved with graphs that link seekers and jobs for promoted channels, and a constrained variant of EBR for organic channels. We also presented a novel GPU-based exhaustive search system, surpassing inverted index-based systems in performance. 

One potential avenue for future exploration is to run a split architecture, relying on term-matching for navigational queries and a vector search system for everything else. This can allow us to optimize the entire system, for example distilling the knowledge of a higher capacity ranking model into a two-tower retrieval model~\cite{hinton2015distilling}.
Enhancing our learning to retrieve models with LLMs fine-tuned for retrieval and online inference is another area of improvement~\cite{wang2022text, wang2023improving}. This would not only improve query and document understanding, but also enhance the system's multilingual capabilities. To improve performance on logographic systems, we can incorporate the multi-level query idea from~\cite{li2021embedding}.
Relevance results could be further improved by utilizing more real-time and short-term personalized signals and social graph features in the retrieval models~\cite{huang2020embedding, li2021embedding}.



\bibliographystyle{ACM-Reference-Format}
\bibliography{li_kdd}
\clearpage
\section*{Appendix}

\subsection*{Acknowledgments}
We thank Ping Li for his insightful advice on quantization, and many talented scientists and engineers at LinkedIn for their help and feedback in this work, . 
  
\subsection*{Optimizing System on GPU}
In this section, we share the detail on our optimization strategies to improve the performance of our hybrid system on GPU.

\textbf{TBR.} 
TBR has two concepts - clauses and attributes. For example, if a document has the following information: $\lbrace geo: [123], skill: [456, 789]\rbrace$, then we say there are 2 clauses, where clause 1 (geo) has 1 attribute, and clause 2 (skill) has 2 attributes. We make two assumptions when storing these information in GPU memory:
\begin{compact_enum}
\item Each document has the same number of clauses 
\item Each document has a fixed amount maximum number of attributes (over all clauses)
\end{compact_enum}

For example, let's say we want to store the following information:
\begin{table}[h]
\small
\vspace{-16pt}
\begin{tabular}{l|l|l}
\hline
  & geo & skill  \\ \hline \hline
doc 1 & 934, 2934 & 945, 342, 3112 \\ \hline
doc 2 & 129 & 9342, 234 \\ \hline
\end{tabular}
\label{table:filters}
\vspace{-12pt}
\end{table}

It will be stored as a matrix (assuming maxNumAttr = 5):
\begin{table}[h]
\small
\vspace{-10pt}
\begin{tabular}{l|l}
\hline
 attributes & offsets  \\ \hline \hline
 934, 2934,  945,  342, 3112 & 0, 2, 5 \\ \hline
 129, 9342,  234,    0,    0 & 0, 1, 3 \\ \hline
\end{tabular}
\label{table:offsets}
\vspace{-12pt}
\end{table}

``Offsets'' is used to record ``begin'' and ``end'' position of each clause. When numAttr < maxNumAttr, we can simply pad with zero, or we can more sophisticated data structures such as linked lists to save the memory space.

When presented with a query containing both geo and skill elements, we can  transform it into Conjunctive Normal Form. We check whether the geo query overlaps with geo attributes of docs and whether the skill query aligns with skill attributes. If both conditions are met, it constitutes a match. This can be done through matrix multiplication. To perform the checking more efficiently, we can sort attributes in advance. Once they are sorted, we can check if there is an intersection between 2 vectors using an efficient \textcolor{blue}{\href{https://takeuforward.org/data-structure/intersection-of-two-sorted-arrays/}{two pointer approach}}. 

\textbf{Quantization.} 
We apply a quantization technique similar to ``OP+ORP''~\cite{li2023oporp} to accelerate EBR. The difference is that we use a simpler scoring function:
\begin{equation}
\sum_{j=1}^{k}{1\lbrace \textrm{sign}(x_j)==\textrm{sign}(y_j) \rbrace}
\end{equation}

Note that we do not use quantization to directly replace embeddings. Instead, it is used as a ``pre-selection'' step, so it does not have to be very accurate. We choose this because it can be implemented efficiently by:
{\small
\begin{verbatim}
     // x, y are int64_t arrays
     int quantScore = 0;
     for (int i = 0; i < numBits / 64; i++) {
         // (1,1) -> 1, (0,0) -> 1, (1,0) -> 0, (0,1) -> 0
         int64_t z = ~(x[i] ^ y[i]);
         // count ``number of ones'' in z 
         quantScore += __popcll(z); 
     }
     return quantScore;
\end{verbatim}
}

There are a few hyper-parameters needs to be tuned, including
\begin{compact_enum}
\item \textit{numBits}: number of quantization bits. We choose 512 bits.
\item \textit{quantK}: number of maximum documents selected by quantization stage. We choose $200 \times k$, where $k$ is the expected number of docs to be returned by the retrieval system.
\end{compact_enum}

We use int64 to store quantized bits. Each int64 can store 64 bits. Quantized embeddings are stored as a dense matrix in GPUs.

\textbf{EBR.}
In the EBR step, we simply do the inner-product of query embedding with doc embeddings, and retrieve top $k$. Note that we do not use cublas APIs such as cublasGemmEx because they do not provide a ``skip rows'' functionality. (For example, after the quantization step, maybe only 100k out of millions of docs needs to be scored. cublasGemmEx don’t have the ability to ``only score those 100k docs''.) For this reason, we write our own CUDA kernel to implement inner-products.

\textbf{Message Passing between Layers.}
A mechanism is needed to pass eligible items from TBR -> Quant, and Quant -> EBR. To do that, we have a data structure called messenger used to pass information. A messenger contains rowId, and other meta data including score and batchId (we will discuss batch soon). For example, if there are 10 docs in the pool, and TBR only selects 3 docs (whose rowIds are 2, 5, 9), then TBR will output an array of messengers like this: 
\vspace{-2pt}
\begin{gather*}
\textrm{[(rowId: 2, metaData), (rowId: 5, metaData), (rowId: 9, metaData)]}
\vspace{-6pt}
\end{gather*} 
This will be the input to the Quant step. This way Quant knows that it only needs to process row 2, 5, and 9.

\textbf{Storing Dense Matrix.}
Recall that embeddings, quantized embeddings, and TBR offsets and attributes are stored as separate dense matrices in GPU memory. We use \textcolor{blue}{\href{https://en.wikipedia.org/wiki/Row-\_and\_column-major\_order}{column major}} to store TBR and quantized embeddings matrices because such arrangement is more efficient in GPU. The table below compares a $(20M \times 100) \times (100 \times 1)$ matrix-vector multiplication. Pass-rate indicates how many rows are needed for scoring. For example, 10\% means only 2M out of 20M rows are scored.

\begin{table}[h]
\small
\vspace{-6pt}
\begin{tabular}{l|l|l}
\hline
 Pass-rate & Row-major latency & Col-major latency  \\ \hline \hline
100\%  & 23.5ms  & 5.7ms \\ \hline
50\%   & 12.8ms  & 5.3ms  \\ \hline
20\%   & 6.5ms & 5.2ms  \\ \hline
10\%   & 3.0ms & 4.3ms  \\ \hline
5\%    & 1.3ms & 2.4ms  \\ \hline
2\%    & 0.6ms & 1.2ms  \\ \hline
1\%    & 0.2ms & 0.8ms  \\ \hline
\end{tabular}
\label{table:col_major}
\vspace{-12pt}
\end{table}

As we can see, col-major leads to better latency in high-pass-rate and row-major leads to better latency in low-pass-rate. We care more about high-pass-rate scenarios, so using col-major is our choice.

For the embedding matrix, we use a special data structure that is a mix of row major and column major, to store them in continuous memory space.
The data structure is designed this way because we can leverage a special data structure called  \textcolor{blue}{\href{https://learn.microsoft.com/en-us/windows/win32/numerics\_h/float4-structure}{float4}}. It allows you to access 4 floating points at once. This optimization gives us 25\% speed up on embedding operations.

\textbf{Fast Top K Selection.} 
To select topK from a list of docs, the most naive way is to sort it first and then select top k docs. Instead, we apply the following trick when performing top K selection.

For example, if we want to find the top $k$ from an array.
First, we define bucketizedScore = (int)(score * 2) + 2,
where 2 = bucketGranularity that we set to 100 in practice. Here we use 2 as it's easier to explain. After bucketization, each element falls into a bucket. Second, we count the number of items in each bucket. Then we start backwards from the bucket corresponding to the largest number, and check at which bucket we can gather slightly more than 4 items. Let's assume this is bucket $l$. We then create a sub-array with elements whose bucketized score >= $l$. Then we sort this sub-array and get top $k$.

Note that this algorithm requires to know the upper and lower bounds of the scores in advance. We conducted an experiment comparing the naive method (using sorting) with our algorithm for selecting 2,000 items from a pool of 1,000,000 candidates, and we see the following improvement:

\begin{table}[h]
\small
\vspace{-6pt}
\begin{tabular}{l|l}
\hline
Naive (sort) & Our method  \\ \hline \hline
1.91ms  & 0.38ms \\ \hline
\end{tabular}
\label{table:bucket_sort}
\vspace{-6pt}
\end{table}

\textbf{Batch Inference.}
In order to increase GPU throughput, we implement a mechanism that bundles a few requests in a batch. This way we only need to scan through the TBR attributes and embeddings once. To implement this, we simply need to add a batchId field in messenger. For example, if you process two queries in a batch, and query 1 has rows $\{1, 2, 5\}$ as eligible items, query 2 has rows $\{3, 5, 9\}$ as eligible items, we will generate the following messenger array:
{\small
\begin{verbatim}
    [(rowId=1, batchId=0), (rowId=2, batchId=0), (rowId=3, 
    batchId=1), (rowId=5, batchId=0), (rowId=5, batchId=1), 
    (rowId=9, batchId=1)]
\end{verbatim}
}

The table below shows the throughput improvements by using batch inference. (numDocs = 15M). Note that using batch is a tradeoff between QPS and latency. For example, let’s say we have 4 requests in total, with batch=1, it may take 5ms + 5m + 5ms + 5ms = 20ms, so the QPS is 1000 / (20 / ) = 200, and the latency is 5ms. With batch = 2, it may take 7.5ms + 7.5ms = 15ms. In this case QPS = 1000 / (15 / 4) = 266.6, and the latency = 7.5ms. 

\begin{table}[h]
\small
\vspace{-6pt}
\begin{tabular}{l|l|l}
\hline
 Batch Size & QPS & Latency  \\ \hline \hline
1  & 570  & 3.2ms \\ \hline
2   & 684  & 4.6ms  \\ \hline
4   & 783 & 6.9ms  \\ \hline
8    & 910 & 10.8ms  \\ \hline
16   & 1101 & 17.2ms  \\ \hline
\end{tabular}
\label{table:Batch_inference}
\vspace{-6pt}
\end{table}

\textbf{Memory Pre-Allocation.}
Finally, to ensure our application will never encounter an OOM issue, all the memory is pre-allocated in the constructor.

\end{document}